\documentclass[9pt,twocolumn]{extarticle}
\usepackage{authblk}
\usepackage[resetlabels,labeled]{multibib}
\newcites{S}{References}

\usepackage[utf8]{inputenc}
\usepackage[english]{babel}
\usepackage{amsmath,amsfonts,amssymb}
\usepackage{mathpazo} 
\usepackage[scaled]{helvet}
\usepackage[T1]{fontenc}
\usepackage{url}
\usepackage[colorlinks=true, allcolors=blue]{hyperref}

\usepackage{graphicx,xcolor}
\usepackage{colortbl}
\usepackage{booktabs}
\usepackage{algorithm}
\usepackage[noend]{algpseudocode}
\usepackage{changepage}
\usepackage[left=48pt,%
                right=42pt,%
                top=46pt,%
                bottom=60pt,%
                headheight=15pt,%
                headsep=10pt,%
                letterpaper,twoside]{geometry}%
\usepackage[labelfont={bf,sf},%
                labelsep=period,%
                figurename=Fig.,%
                singlelinecheck=off,%
                justification=RaggedRight]{caption}
\usepackage{amsmath,amssymb,graphicx}

\usepackage{physics}
\DeclareMathAlphabet\mathbfcal{OMS}{cmsy}{b}{n}

\title{High resolution mid-infrared optical coherence tomography with kHz line rate}

\author[1,2,*]{Niels M. Israelsen}
\author[1]{Peter John Rodrigo}
\author[1,2]{Christian R. Petersen}
\author[1]{Getinet Woyessa}
\author[1]{Rasmus E. Hansen}
\author[1,3]{Peter Tidemand-Lichtenberg}
\author[1,3]{Christian Pedersen}
\author[1,2,4]{Ole Bang}

\affil[1]{DTU Fotonik, Technical University of Denmark, 2800 Kgs, Lyngby, Denmark}
\affil[2]{NORBLIS IVS, Virumgade 35 D, Virum 2830, Denmark}
\affil[3]{NLIR ApS, Hirsemarken 1, Farum, 3520, Denmark}
\affil[4]{NKT Photonics, Blokken 84, 3460 Birkeroed, Denmark}
\affil[*]{Corresponding author: nikr@fotonik.dtu.com}

\begin{document}

\maketitle

\begin{abstract}
 We report on Mid-infrared (MIR) OCT at 4 $\mu$m based on collinear sum-frequency upconversion and promote the A-scan scan rate to 3 kHz. We demonstrate the increased imaging speed for two spectral realizations, one providing an axial resolution of 8.6 µm, and one providing a record axial resolution of 5.8 µm. Image performance is evaluated by sub-surface micro-mapping of a plastic glove and real-time monitoring of CO$_2$ in parallel with OCT imaging.
\end{abstract}
\section{Introduction}
Mid-infrared optical coherence tomography (MIR OCT) was first introduced in 2007 by Colin et al \cite{colley2007mid}. It is however only recently that the technology has become mature enough to harness the deep penetration ability when imaging turbid media. In one approach a supercontinuum (SC) source is combined with a slow pyroelectric array detector \cite{zorin2018mid} where a second is based on frequency entanglement via spontaneous parametric downconversion \cite{vanselow2020frequency}. A third technology was realized by the Authors and is based on an SC source combined with sum-frequency upconversion \cite{israelsen2019real}. To date, this solution still holds the record in axial resolution and imaging speed of 8.6 µm and 0.3 kHz, respectively. 

In this paper we report on an improvement of the sum-frequency upconversion detection scheme that provides a tenfold increase in imaging speed to a 3 kHz line rate. Additionally, by introducing an improvement of the optical setup, we present two spectral bandwidth realizations, one providing an 8.6 µm axial resolution, and one providing a record 5.8 µm axial resolution. We evaluate the imaging performance for the two bandwidths and we characterize the noise, sensitivity and axial resolution. As specific demonstration cases we image a nitrile plastic glove and provide a temporal demonstration of real-time monitoring of CO$_2$ in parallel with OCT imaging.

\section{Methods and materials}
\begin{figure}[htbp]
\centering
\fbox{\includegraphics[width=0.99\linewidth]{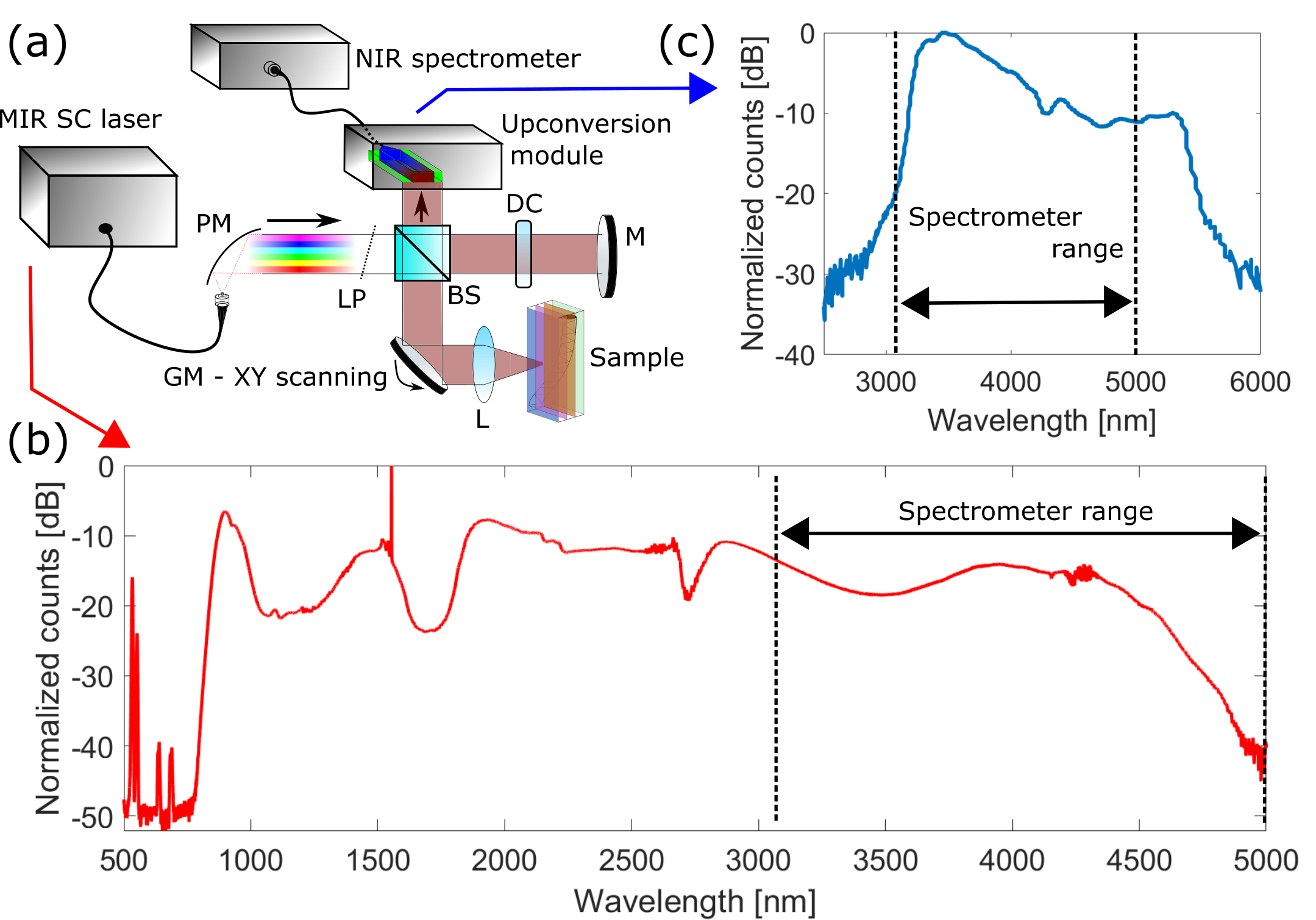}}
\caption{4 µm MIR OCT setup: (a) Schematic of OCT setup. Abbreviations: PM, parabolic mirror, LP, longpass filter, BS, beamsplitter, DC, dispersion compensating component, M, silver coated mirror, GM, galvanometric scanning mirrors, and L, lens. (b) Spectrum of the supercontinuum source and (c) spectrum of the upconversion module using a glow bar. Double arrow marks the range of the spectrometer.}
\label{Setup}
\end{figure}

The MIR OCT system presented in this work is an improved version of the first system used in \cite{israelsen2019real}. It exploits the latest development in MIR SC and broadband sum-frequency upconversion technology. A schematic of the OCT system is presented in Fig.~\ref{Setup}(a). A custom-made MIR SC source based on a 1.55 µm master-oscillator power amplifier (MOPA) pump laser with a pulse duration of 1 ns and repetition rate of 2 MHz, is pumping a ZBLAN fiber, which provides a continuous spectrum from 0.8 to 5 µm 
as shown in Fig.~\ref{Setup}(b). 
The light beam is subsequently long-pass (LP) filtered to remove unexploited power in the near-infrared (NIR) range with either a 3.5 µm or a 2.4 µm LP filter yielding the bandwidths BW1 (3.5 - 5.0 µm) or BW2 (2.4 - 4.9 µm), respectively. By using BW2 instead of BW1
we allow for utilization of more of the full detection range for the combined spectrometer and upconversion module (3.2 - 5.1 µm). After the spectral selection, the light beam enters the Michelson interferometer with a reference and a sample arm. 
In the sample arm the sample is exposed to either 18 mW (BW1) or 52 mW (BW2). The full detection range (3.2 - 5.1 µm) corresponds to 22 mW. The sample is scanned with a lateral resolution of $\sim 15$µm \cite{israelsen2019real}.

After recombination of the reference and sample beam, the mixed beam is focused into the upconversion module where upconversion from MIR wavelengths to NIR wavelengths is carried out based on sum-frequency generation. The module is composed of a 1064 nm laser cavity with a periodically poled lithium-niobate crystal. In contrast to the recent upconversion system applied in \cite{barh2017ultra,israelsen2019real}, where the poling period is constant and the broadband phase-matching is accomplished by angularly distributing the light beam by a tight beam focus, the broadband phase-matching in this setup is achieved collinearly \cite{friis2019upconversion}. This is done by exploiting a chirped periodically poled crystal with the poling period linearly decreasing from 23.5 µm to 22 µm from input to output over a 20 mm length. By mixing the 1064 nm CW laser with the broad MIR spectrum, the MIR wavelengths are converted to NIR wavelengths with spectral components converted separately in space along the propagation axis dictated by the phase-matching condition $k_{MIR}+k_{pump}+k_\Lambda=k_{NIR}$, where the respective k-vectors are those of the MIR wavelength, the 1064 nm mixing laser, the poling periodicity $\Lambda$, and the NIR wavelength. The spectral range (or bandwidth) of the upconversion module, measured with a glow bar as light source, is shown in Fig~\ref{Setup}(c). Finally, the upconverted light is guided by a single mode fiber to a commercial NIR spectrometer (Wasatch Photonics) with 4096 elements covering the wavelength range of 796-879 nm with a maximum line rate of 40 kHz. An electronic gain setting was selected to maximize the effective dynamic range. In addition to the MIR OCT system, a customized NIR OCT system designed for dermatological imaging \cite{israelsen2018value} was utilized for comparison. To construct OCT images, normalization, k-space re-sampling, and general dispersion compensation were carried out \cite{israelsen2018value,bradu2018recovering}.
\section{Results}

\subsection{Noise characterization}
\begin{figure}[h!]
\centering
\fbox{\includegraphics[width=0.99\linewidth]{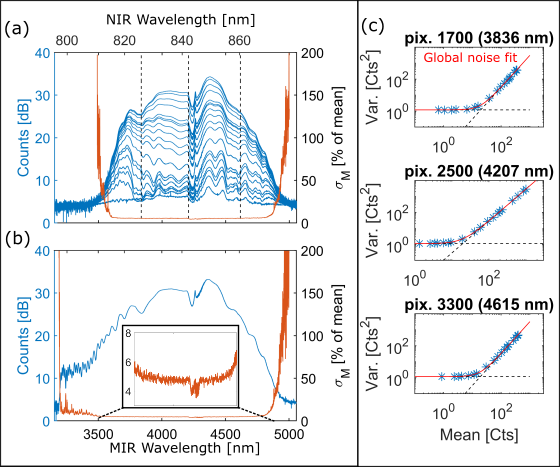}}
\caption{Spectral noise evaluation. (a), (b) dark count subtracted average spectra and relative standard deviation noise $\sigma_M$ (RHS y-axis) using BW1 and BW2, respectively. Inset in (b) shows a noise base line of 5\%. (c) noise evaluation based on the series of averaged spectra in (a) for three selected spectrometer pixels marked with vertical dashed lines. The global variance noise fit (red line), the read/dark noise, $\sigma_{r+d}^2$, and excess noise, $\sigma_{ex}^2$ (black dashed lines), are included. Spectrometer integration time setting was 300 µs.}
\label{noise}
\end{figure}

The first step in evaluating the MIR OCT system was to characterize the noise properties. The raw spectra with dark count subtraction is presented for BW1 in Fig.~\ref{noise}(a) for different reference arm power levels (sample arm blocked), and for BW2 in Fig.~\ref{noise}(b) for the maximum reference power level. Each spectrum is an average of 500 sequentially recorded single spectra with a spectrometer integration time of 300 µs. This tenfold decrease in integration time, compared to our previous report \cite{israelsen2019real}, is possible due to the improved collinear upconversion efficiency which significantly increases the strength of the interferometric OCT signal. 

\indent In OCT the noise is comprised of three components: the read (digitization and read-out) and dark noise $\sigma_{r+d}$ dictated by the detector, the shot noise $\sigma_{SN}$ dictated by the photon statistics, and finally the excess noise, $\sigma_{ex}$, typically stemming from the spectral power fluctuations of the light source, which are significant when using an SC source. Together they present the total noise fluctuations seen in a detected spectrum described in variance as $\sigma^2_{tot}= \sigma_{r+d}^2+\sigma_{SN}^2+\sigma_{ex}^2$. When evaluating the single pixel noise we calculate the relative total noise (standard deviation) $\sigma_{M}$, with M designating a pulse train of M pulses within the spectrometer integration time. We find for the maximum signal (spectrum) a $\sigma_{M}$ in the range of $4-6\%$ (red graphs, Fig~\ref{noise}(a) and \ref{noise}(b)) where the best NIR SC sources for OCT show levels of around $1\%$ \cite{jensen2019noise}. As we shall see, for these high signal levels the noise is solely dominated by excess noise of the SC source.  The $100\%$ boundaries for $\sigma_M$ in Fig.~\ref{noise}(a) (largest power level) and (b) express the bandwidth of BW1 and BW2 to 3.5-5.0 µm and 3.2-4.9 µm, respectively. We note that the dips in the spectra of Fig.~\ref{noise}(a) and (b) at $\sim 4.3$ µm wavelength is a result of the ambient atmospheric CO$_2$ absorption.

The next step is to decompose the three noise components of $\sigma^2_{tot}$. The term $\sigma_{r+d}^2$ is expected to be independent of laser power, where $\sigma_{SN}^2$ obeys Poissonian statistics and thus scales linearly with the mean power registered by the spectrometer. $\sigma_{ex}^2$ scales with the mean power squared. It depends on the pulse repetition rate and the pulse-to-pulse relative intensity noise (RIN) of the SC source determined by the nonlinear dynamics of the SC generation process \cite{kwarkye2020amplifier}. As our SC source is using long pump pulses, the SC generation is based on modulational instability amplifying quantum noise and is thus inherently noisy \cite{Dud06Rev}. We therefore do not expect to have access to the shot-noise limited regime \cite{jensen2020shot}. %
This is confirmed in Fig.~\ref{noise}(c) where single pixel signals of the spectral series of Fig.~\ref{noise}(a) are evaluated through the variance vs. mean signal. Fitting data sets by the total noise model with a, b, and c parameters acting as weighting of the three noise terms, we find that within the spectrometer dynamic range the noise is entirely dominated by either $\sigma_{r+d}^2$ for low signal levels, or $\sigma_{ex}^2$ for high signal levels. The detector noise is found to render $\sigma_{r+d}\sim 1$ count. An optimum operation regime is found at $\sim 20 $ counts where $\sigma_{r+d}^2\sim\sigma_{ex}^2$.

\subsection{OCT performance}
\begin{figure}[h!]
\centering
\fbox{\includegraphics[width=0.99\linewidth]{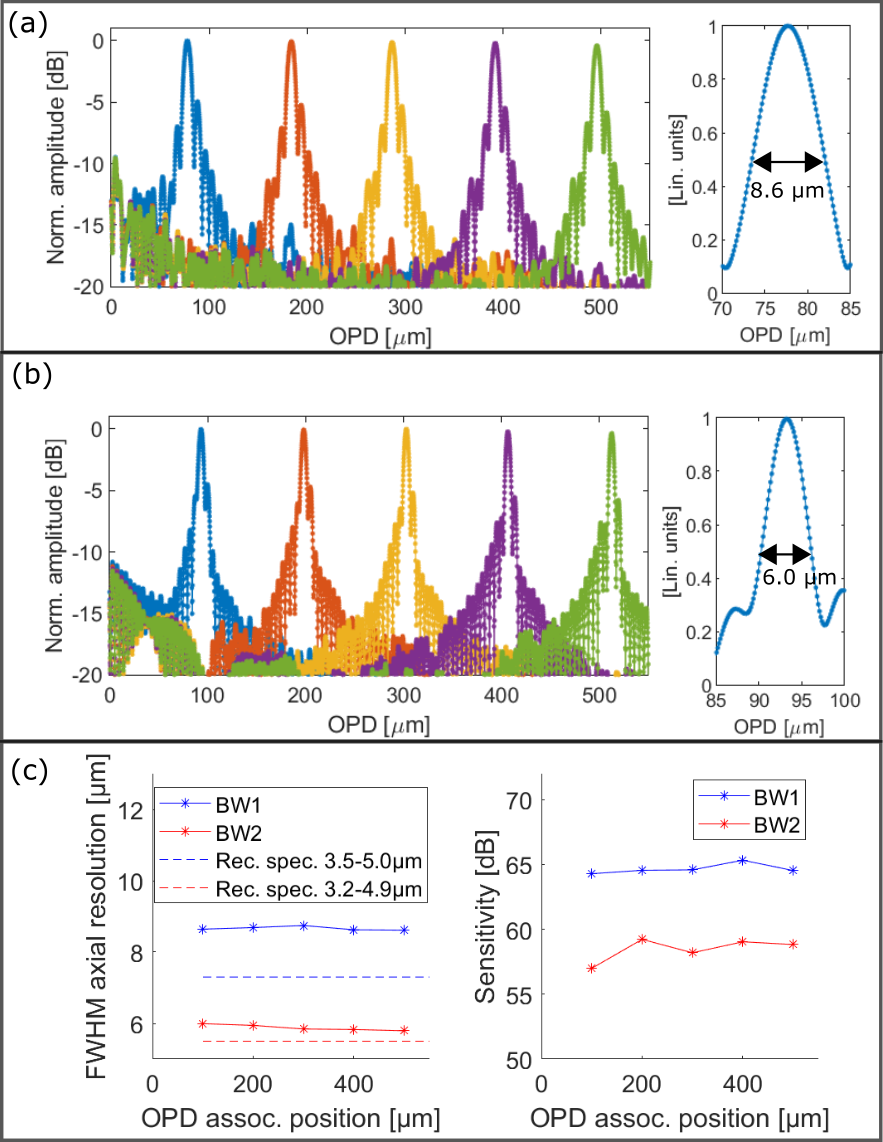}}
\caption{MIR OCT axial resolution and sensitivity for a 3 kHz line rate. (a), (b) A-scans and a zoom-in for the shortest optical path difference (OPD) when using BW1 and BW2, respectively. (c) Summary of the axial resolution and sensitivity evaluation.}
\label{Fig3-OCT}
\end{figure}

The next step was to evaluate the axial resolution and sensitivity performance of the MIR OCT system operated at a 3 kHz line rate. A mirror was inserted as a sample and a controlled attenuation of the signal from this was introduced to avoid detector saturation. In doing so, the point-spread-function limited A-scan was generated. A vital part of this is careful re-sampling from wavelength space to wavenumber space and dispersion compensation to remove the residual dispersion mismatch between reference and sample arm \cite{bradu2018recovering}. Zero-padding was done prior to the Fourier transform in order to mitigate any sampling influence on resolution and sensitivity. 
Examples of A-scans for BW1 and BW2 are presented in Fig.~\ref{Fig3-OCT}(a) and (b), respectively, producing a signal 6 dB roll-off at 2.1 mm (not in figure). The respective zoom-ins of the shortest optical path difference (OPD) A-scan and associated axial resolution is marked. A summary of the associated A-scan resolutions is found in Fig.~\ref{Fig3-OCT}(c) (left). These are extracted by full-width half-maximum (FWHM) readings. For BW1 an axial resolution of 8.6-8.7 µm was found, matching that reported previously \cite{israelsen2019real}. For BW2, we find an improved axial resolution of 5.8-6.0 µm, which to our knowledge is a record in $\sim4$µm center wavelength OCT.
The theoretical resolutions of rectangularly shaped spectra defined by BW1 and BW2 (given by the FWHM of a sinc-shaped central lobe after Fourier transform) are marked by dashed lines in Fig.~\ref{Fig3-OCT}(c). As the read-off FWHM is strongly dependent on the spectral shape this theoretical comparison primarily reveals deviation in energy distribution between the perfect rectangular spectrum and a irregularly-shaped experimental spectrum.

Sensitivity values associated with the A-scans of Fig.~\ref{Fig3-OCT}(a) and (b) are presented in Fig.~\ref{Fig3-OCT}(c) (right). BW1 and BW2 demonstrates sensitivities of $\sim$65 dB and $\sim$59 dB, respectively, the difference being a result of the different reference signals applied. These values resemble that reported previously when using non-collinear upconversion \cite{israelsen2019real}.


\subsection{Spatial and temporal imaging}
To demonstrate imaging capabilities, results from two imaging sessions are presented. The first is of a cut-out of a nitrile plastic glove (Dermatril) from Honeywell, USA, certified for laboratory use. The glove represents a highly scattering material in the NIR where MIR OCT is expected to provide better penetration. This fact is documented in Fig.~\ref{Fig4-OCT}(a)-(c). B-scans of the glove produced by applying BW1 and BW2 are depicted in Fig.~\ref{Fig4-OCT}(a) and (b), respectively, where penetration through the two layers of plastic is manifested in imaging the underlying metal platform on which the glove rests. In Fig.~\ref{Fig4-OCT}(a) BW1 enables visualization of the backside of the second layer where it is only clearly visible in a single point for BW2 in Fig.~\ref{Fig4-OCT}(b), due to the reduced sensitivity. As seen from Fig.~\ref{Fig4-OCT}(c), our NIR OCT system does not penetrate even a single plastic layer and from the surface a signal extinction tail is seen, similar to what is observed when multiple scattering comprises the majority of the depth signal \cite{israelsen2019real}. For this reason, the NIR OCT scan is taken across the edge of the cut-off in a "staircase" manner in order to visualize both plastic layers and the metal platform. 

The second demonstration is performed by spectroscopically monitoring CO$_2$ deviations in the air above an aluminium plate. A photo of the sample setup is given in Fig.~\ref{Fig4-OCT}(d). A voluntary person blows into a plastic tube extending down to the plate, as illustrated in Fig.~\ref{Fig4-OCT}(e), where the tube termination is arranged so no air flow enters the interferometer. During this exercise, A-scans are recorded at a fixed single point on the metal plate, the position indicated in the C-scan in Fig.~\ref{Fig4-OCT}(f). In that way, a temporal B-scan is constructed as depicted in Fig.~\ref{Fig4-OCT}(g). Fig.~\ref{Fig4-OCT}(h) shows the spectral decomposition of the corresponding interferograms into two sub-windows: one being a coarse CO$_2$ window at 4.27 µm with a FWHM (Gaussian) width of 295 nm (red dashed line) and one the entire spectrum (BW1) subtracted by the CO$_2$ window (black dashed line). The integrated signals in the two spectral windows are then evaluated during a single human exhalation in a time slot of 10 seconds as presented in Fig.~\ref{Fig4-OCT}(i). The ratio between the two windowed signals is depicted in Fig.~\ref{Fig4-OCT}(j) (blue data points). First a mechanical instability at 1.2s is seen as a sharp dip indicating initiation of the exhalation and thereby translation of the air column in the tube. Secondly, a significant drop to a relative signal intensity of 0.92 is seen lasting four seconds before the initial condition is regained. The four seconds corresponds to the exhalation time of our test person manifested in a smooth spectral transition. A retake of the experiment with two exhalations is also presented in Fig.~\ref{Fig4-OCT}(j) (orange data points). It is important to point out that each data point in Fig.~\ref{Fig4-OCT}(i) and (j), is an average of 100 spectra, each spectrum measured over a 300 µs integration. The finer temporal resolution is indicated by the inset of Fig.~\ref{Fig4-OCT}(j) which also marks the spectral data of Fig.~\ref{Fig4-OCT}(h) as a red data point.

\section{Discussion and conclusion}
Going from a non-collinear to a collinear upconversion provides unequivocally a significant advancement in the MIR OCT performance. The advancement was demonstrated for a 300 µs spectrometer integration time and for two different source bandwidths, BW1 and BW2.
BW1 and BW2 were for the most of the detector dynamic range dominated by excess noise; not surprising when utilizing a modulation instability based SC source. From these the best case axial resolutions of 8.6 µm and 5.8 µm were produced, the latter a new record in MIR to the best of our knowledge. 

\begin{figure}[h!]
\centering
\fbox{\includegraphics[width=0.99\linewidth]{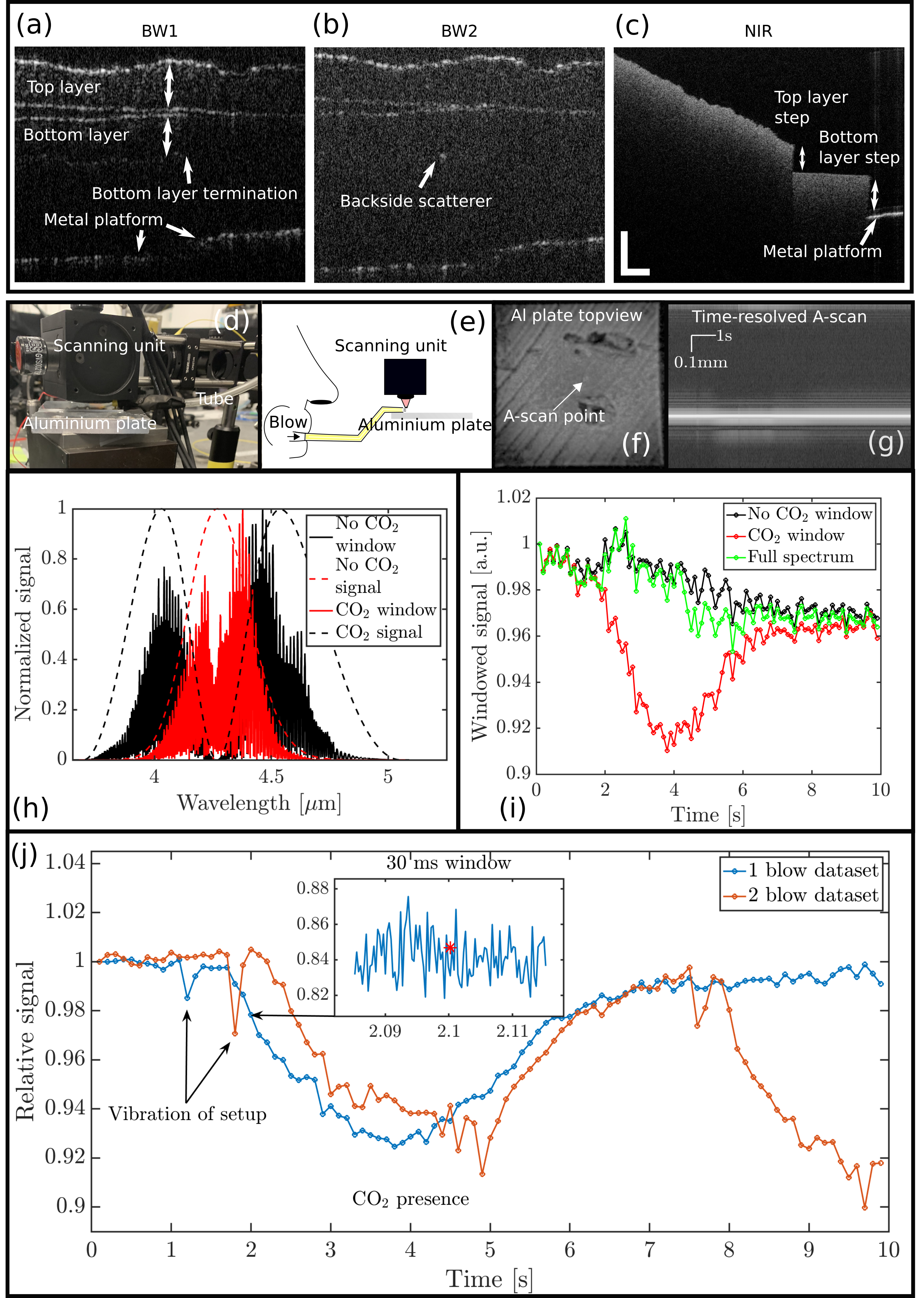}}
\caption{\textbf{Imaging of plastic nitrile glove:}  (a)-(c) B-scans of the two layers of the glove for BW1, BW2 and 1300 nm OCT, respectively. Scale bars mark globally 200 µm in air. \textbf{Spectroscopic monitoring of CO$_2$ exhalation}: (d) Sample setup, (e) simplified sketch of this. (f) Location of monitoring as a C-scan and (g) temporal B-scan at this location. (h) The two spectral windows chosen for the spectroscopic interrogation. (i) Example of 10s measurement of the signals for the two spectroscopic windows. (j) Ratio between the two signal windows for both a single exhalation and two subsequent exhalations.}
\label{Fig4-OCT}
\end{figure}

An image demonstration was performed on a plastic glove (Fig.~\ref{Fig4-OCT}(a)-(c)) with the advantage in penetration seen when comparing the NIR OCT image with MIR OCT images of BW1 and BW2. What is noteworthy is the difference in Fig.~\ref{Fig4-OCT}(a) and (b) where the backside of the second plastic layer is visible for (a), but this signal is entirely buried in noise for (b) apart from in a single point. Also the difference in the general background noise can be recognized. This example underlines the important outcome of the sensitivity performance when imaging a curving and rough surface, as a glove, where the $\sim6$ dB sensitivity difference between BW1 and BW2 (Fig.~\ref{Fig3-OCT}(c)) is decisive when tracking such a sub-surface interface. Careful consideration is thus needed when prioritizing between axial resolution and sensitivity in applying our system. 

A simple temporal imaging demonstration was performed as a proof-of-principle to highlight the system's unique ability to spectroscopically monitor CO$_2$ flow in parallel with OCT imaging. With a temporal resolution of 300 µs we show that this record imaging speed can follow real-time gas-flow fluctuations, but is however susceptible to chromatic effects from mechanical vibrations resulting in misleading spectroscopic signals as indicated in Fig.~\ref{Fig4-OCT}(j). This means that it is essential to have the sample setup absolutely fixed during spectroscopic measurements in order to mitigate error sources spurred by  spatial chromatism.

In conclusion, we have presented a new MIR OCT scheme based on collinear sum-freqeuncy upconversion introducing kHz line rate MIR OCT imaging and an axial resolution below 6 µm. With this, we have reached a speed where real-time gas flow tracing can be monitored in the setting of OCT. 

\section{Disclosures} The authors declare no conflicts of interest.
\section{Funding} We acknowledge support by Innovation Fund Denmark under the project J. No. 4107-00011A (ShapeOCT).
\clearpage

\bibliographystyle{unsrt}
\bibliography{main}

\end{document}